\documentclass[conference]{IEEEtran}
\IEEEoverridecommandlockouts
\usepackage{cite}
\usepackage{amsmath,amssymb,amsfonts}
\usepackage{algorithm}
\PassOptionsToPackage{noend}{algpseudocode}% comment out if want end's to show

\PassOptionsToPackage{noend}{algpseudocode}% comment out if want end's to show

\usepackage{graphicx}
\usepackage{textcomp}
\usepackage{xcolor}
\usepackage{algpseudocode}
\usepackage{subfig}

\usepackage{url}
\usepackage{color,soul}
\usepackage{tikz}
\usepackage{makecell}
\usepackage{multirow}
\def\BibTeX{{\rm B\kern-.05em{\sc i\kern-.025em b}\kern-.08em
    T\kern-.1667em\lower.7ex\hbox{E}\kern-.125emX}}
\begin{document}

\title{Leveraging 5G Physical Layer Monitoring for Adaptive Remote Rendering in XR Applications}
% \title{XR Edge Rendering architecture for holo-conferencing and E2E performance assessment}
% \\

\author{
\IEEEauthorblockN{Inhar Yeregui\IEEEauthorrefmark{1}, Daniel Mej\'ias\IEEEauthorrefmark{1}, Mikel Zorrilla, Roberto Viola}
\IEEEauthorblockA{\textit{Fundaci\'on Vicomtech} \\
\textit{Basque Research and Technology Alliance}\\
San Sebasti\'an, Spain\\
iyeregui@vicomtech.org}
\IEEEauthorblockA{\IEEEauthorrefmark{1}PhD Candidate at UPV/EHU}
\and
\IEEEauthorblockN{Jasone Astorga, Eduardo Jacob}
\IEEEauthorblockA{\textit{Department of Communications Engineering} \\
\textit{University of the Basque Country}\\
Bilbao, Spain\\
\{jasone.astorga, eduardo.jacob\}@ehu.eus}

% \and
% \and
% \and
}

\maketitle

\IEEEoverridecommandlockouts
\IEEEpubid{\begin{minipage}{\textwidth}\ \\\\\\\\\\[12pt]\centering
979-8-3315-1998-8/25/\$31.00 ©2025 IEEE
\end{minipage}}
\begin{abstract}
As immersive eXtended Reality (XR) applications demand substantial network resources, understanding their interaction with 5G networks becomes crucial to improve them. This paper investigates the role of 5G physical-layer monitoring to manage and enhance the remote rendering of XR content dynamically. By observing network metrics directly from the physical layer, we propose a system to adapt streaming parameters such as bitrate, framerate, and resolution in real time based on available network capacity. Using theoretical formulas to estimate maximum data rate, our approach evaluates network resource availability, enabling the renderer to self-adjust media content representation. This is critical for providing consistent and smooth XR experiences to users, especially as network conditions fluctuate.
% Our findings suggest that physical-layer monitoring offers valuable insights into network capacity, which can be leveraged to increase the Quality of Service (QoS) in real time. This approach has the potential to elevate user experiences in remote-rendered XR applications by ensuring that resource allocation aligns closely with network availability.
Our findings suggest that physical-layer monitoring offers valuable insights to increase the Quality of Service (QoS) and has the potential to elevate user experience in remote-rendered XR applications.
\end{abstract}


\section{Introduction}

Extended Reality (XR), encompassing Virtual Reality (VR) and Augmented Reality (AR), is revolutionizing how users interact with digital environments. These technologies enable immersive experiences that blend the physical and digital worlds, offering applications in gaming, education, healthcare, and beyond. However, the resource-intensive nature of XR poses significant challenges, particularly when rendering complex content in real time. High-fidelity XR applications require substantial computational power to generate realistic 3D environments and process interactions at low latency. While some XR devices, such as high-end PCs or gaming consoles, can handle local rendering, many XR applications rely on remote rendering—offloading the computational workload to external servers or Multi-access Edge Computing (MEC) nodes. This approach enables lightweight XR devices, such as standalone headsets or mobile devices, to access high-quality experiences without the need for expensive onboard hardware. However, remote rendering introduces a strong dependency on network performance, as the rendered frames must be streamed in real time to the XR device with minimal latency.

Remotely rendered high-quality XR experiences depend on ultralow latency, high bandwidth, and consistent network performance, making the interplay between XR applications and network infrastructure a critical research area \cite{clemm2020toward}. The advent of 5G networks, with their reduced latency capabilities and enhanced bandwidth, has created new opportunities for streaming remote-rendered XR content efficiently \cite{chen20235g}. Despite these advancements, network conditions in 5G environments can still fluctuate due to factors such as user mobility, network congestion, and environmental interferences. These factors can degrade both Quality of Experience (QoE) and the Quality of Service (QoS), manifesting as delays, lower resolution, or disrupted interactions. 

Addressing this challenge requires a deeper understanding of the network behavior and adaptive mechanisms that can respond in real time to changes in network conditions \cite{taleb2022toward}. This paper explores the potential of leveraging 5G physical-layer monitoring to tap into metrics such as signal strength, channel quality, and available bandwidth. By estimating network resource availability, XR streaming systems can dynamically adjust parameters such as bitrate, framerate, and resolution to enhance performance, ensuring a smooth and immersive experience.

% The paper contributes to the state of the art in the following ways:
The contributions of this paper are as follows:

\begin{itemize}

    %\item Design of a modular edge rendering architecture for XR experiences where high computational capability demanding tasks, such as the renderization of complex 3D environments are offloaded to the MEC, to support a heterogeneous variety of end devices. 

    \item Utilizing physical layer monitoring to estimate available network resources: The paper demonstrates how monitoring specific metrics from the 5G network's physical layer, along with theoretical formulas to calculate data rate, can provide an accurate estimate of available network resources. This estimation allows the system to notice changes in connection quality and adjust to maintain a stable user experience in XR environments.

    \item Evaluation of the impact of adaptability on QoS: Through comparing the results to the ones obtained when no adaptability is implemented, this paper evaluates how auto-adaptation based on network resources can enhance the user experience by improving performance and reducing streaming issues in XR experiences. The research provides data on how controlling rendering quality parameters can significantly contribute to consistent QoS in variable connection scenarios.
\end{itemize}

The rest of the paper is structured as follows. Section \ref{sec:RelatedWork} reviews related work on delivering immersive XR experiences over 5G solutions. Section \ref{sec:estimation}, focuses on the proposed physical layer-driven bandwidth estimation method, while its implementation in an auto-adaptive XR Remote Rendering system is described in section \ref{sec:implementation}. Section \ref{sec:evaluation} details the experimental assessment and finally, Section \ref{sec:conclusions} presents conclusions and outlines future work directions.

\section{Related Work}
\label{sec:RelatedWork}

%------------------------------EXTENDED ABSTRACT VERSION-------------------------------------------
% (Extended abstract version) This section highlights recent research on the interplay between Extended Reality (XR) technologies and 5G networks. Studies such as \cite{gapeyenko2023standardization}, \cite{sampath2024enabling}, and \cite{siriwardhana2021survey} address challenges like network latency, bandwidth demands, and Quality of Experience (QoE), emphasizing the importance of adaptive mechanisms in dynamic network environments. Research on 5G architectures, as in \cite{akyildiz2022wireless}, underscores the potential of next-generation networks to mitigate latency issues in XR, highlighting challenges like user mobility. Contributions such as \cite{laha2023ai} and \cite{bojovic2023enhancing} explore AI-driven resource allocation and monitoring mechanisms to optimize streaming performance under fluctuating conditions.

% Despite these advances, existing solutions often overlook XR-specific strategies integrating physical-layer insights for real-time adaptation. This paper fills this gap by leveraging 5G physical-layer metrics to estimate available network resources and adapt XR streaming parameters dynamically, offering a scalable and robust approach to enhance QoE in 5G-enabled XR systems.
%---------------------------------FINAL PAPER VERSION-------------------------------------------
The convergence of XR technologies and 5G networks has become a critical research area in recent years. Several studies \cite{gapeyenko2023standardization},\cite{sampath2024enabling},\cite{siriwardhana2021survey} have explored the unique challenges associated with delivering immersive XR experiences over 5G, focusing on network latency, bandwidth demands, and QoS. These works highlight the importance of developing adaptive mechanisms to address fluctuations in network conditions and ensure seamless user experiences.

Research on 5G networks has emphasized their potential to support low-latency, high-bandwidth applications. Authors in \cite{akyildiz2022wireless} have highlighted how 5G and 6G architectures can be leveraged to mitigate latency issues for XR, enabling interactive and immersive applications. However, this work also identifies challenges such as user mobility, which can disrupt the reliability of XR streaming.

The role of adaptive streaming systems in XR has also been the focus of several studies. Studies such as \cite{laha2023ai} and \cite{liubogoshchev2021adaptive} investigated AI/ML-driven solutions for dynamic resource allocation in XR applications, showcasing the benefits of predictive models for real-time adjustments in bitrate and resolution. Similarly, in \cite{bojovic2023enhancing} a loopback mechanism is proposed to monitor and optimize network performance, demonstrating its ability to enhance QoS under fluctuating conditions.

Other contributions, like \cite{stoica2024xr} have explored how XR-aware next-generation networks, spanning from the physical layer to the applications, enable the optimization and resource allocation to meet the demanding QoS requirements of XR services. 

Despite these advancements, most solutions rely on general optimization frameworks, leaving a gap in XR-specific strategies that integrate physical-layer insights into adaptive streaming.

While the current body of research addresses critical aspects of XR performance over 5G, significant challenges remain. Many existing solutions lack scalability for dense user environments or fail to account for the variability of 5G networks in mobile scenarios. Moreover, few studies have explicitly focused on leveraging physical-layer monitoring to predict resource availability in real time.

This paper builds upon these studies by proposing a novel approach that utilizes physical-layer metrics to enhance XR streaming systems. It focuses on calculating the theoretical maximum data rate of a given system based on 5G wireless networks. By leveraging this data rate estimation, the proposed approach enables XR systems to dynamically adapt streaming parameters in response to real-time network conditions. By bridging the gap between network-level data and real-time adaptation, the proposed method offers a more robust solution for maintaining high QoE in dynamic 5G environments. This contribution addresses the limitations of prior work and paves the way for more reliable and immersive XR experiences.

\section{Physical layer-driven bandwidth estimation}
\label{sec:estimation}

In several 3GPP releases, different network components, e.g., Network Exposure Function (NEF) \cite{kekki2018mec}, are standardized to provide access to network information such as configurations and metrics. Accessing such information enables any application to estimate the current network capacity and improve its behavior. Moreover, \cite{3gpp_ts38_306_v15_2} includes the specifications of the radio access capabilities required for 5G User Equipment (UE) to operate in 5G NR (New Radio) networks. Among others, it describes how to calculate the maximum theoretical data rates for both downlink (DL) and uplink (UL), considering modulation, coding schemes, physical resource blocks (PRBs), and network overhead. The document provides configurations for channels and carriers, including the number of PRBs available for different bandwidths and subcarrier spacing. It also defines requirements for modulation orders and coding efficiency. This specification plays a critical role in ensuring interoperability between devices and networks while guiding manufacturers, operators, and researchers in the development and deployment of 5G technologies.

According to the mentioned specification, the maximum data rate of a 5G network for a user can be calculated using the following equation:

% \[
% \text{Data Rate (in Mbps)} = 10^{-6} \cdot \sum_{j=1}^{J} \left( 
% v_{\text{Layers}}^{(j)} \cdot Q_m^{(j)} \cdot f^{(j)} \cdot R_{\text{max}} \right.
% \]

% \[
% \left. \cdot 
% \frac{N_{\text{PRB}}^{BW(j), \mu} \cdot 12}{T_s^\mu} \cdot (1 - OH^{(j)})
% \right)
% \]
\begin{equation}
\label{deqn_ex1}
\begin{split}
\text{Data Rate (Mbps)} = 10^{-6}\cdot\sum_{j=1}^{J} \bigg( 
v_{\text{Layers}}^{(j)} \cdot Q_m^{(j)} \cdot f^{(j)} \cdot R_{\text{max}} \cdot \\
\frac{N_{\text{PRB}}^{BW(j), \mu} \cdot 12}{T_s^\mu} \cdot (1 - OH^{(j)})
\bigg)
\end{split}
\end{equation}

The parameters presented in the formula are:
\begin{itemize}
    \item \(J\): Number of aggregated component carriers in a band or band combination.
    \item \(v_{\text{Layers}}^{(j)}\): Number of transmission layers for carrier \(j\).
    \item \(Q_m^{(j)}\): Modulation order for carrier \(j\), i.e., bits per symbol. E.g., 2 for Quadrature Phase Shift Keying (QPSK), 4 for 16 Quadrature Amplitude Modulation (16-QAM), etc.
    \item \(f^{(j)}\): The scaling factor. It can take the values 1, 0.8, 0.75 and 0.4.  
    \item \(R_{\text{max}}\): Maximum code rate (e.g., \( R_{\text{max}} = \frac{948}{1024} \)).
    \item \(N_{\text{PRB}}^{BW(j), \mu}\): Number of PRBs available for carrier \(j\) in the given bandwidth \(BW\) and numerology \(\mu\).
    \item \(T_s^\mu\): Symbol duration in seconds, which depends on the subcarrier spacing (e.g., for \(\Delta f = 15 \, \text{kHz}, T_s = 71.4 \, \mu s\)).
    \item \(OH^{(j)}\): Overhead for carrier \(j\), representing the fraction of resources not used for data (e.g., \(OH = 0.14\) for frequency range FR1 for DL).
\end{itemize}

The information derived from the equation can be directly applied to a content streaming system with adaptive bitrate mechanisms. By periodically calculating the maximum theoretical data rate available on the 5G channel, the system can dynamically select the most suitable streaming profile based on current network conditions.
%For instance, under poor channel conditions, the system may opt for a lower resolution (e.g., 720p) and lower bitrate to ensure uninterrupted streaming. Conversely, when the channel quality improves, the system can increase the resolution and bitrate (e.g., 1080p or 4K) to provide a better user experience. This approach enables real-time adaptation to network variability, ensuring optimal resource utilization and maintaining service quality.

\section{Adaptive Remote Rendering for XR}
\label{sec:implementation}

\begin{figure}[t!] 
     \centering
     \includegraphics[width=1\linewidth]{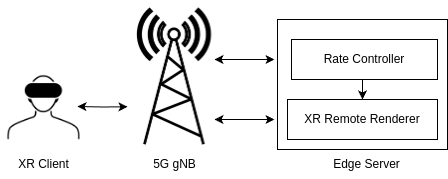}
     \caption{Architecture for auto-adaptive XR Remote Rendering.}
     \label{fig:testbed}
 \end{figure}
 
Figure \ref{fig:testbed} shows the general architecture of an auto-adaptive XR Remote Rendering system. It consists of two main modules: the Rate Controller and the XR Remote Renderer. The Rate Controller is responsible for collecting network metrics and conducting bandwidth estimations that are going to be the input for the XR Remote Renderer to adapt the content streaming. Then, an XR Client receives the rendered media content through the 5G network. 

\begin{figure}[t] 
     \centering
     \includegraphics[width=1\linewidth]{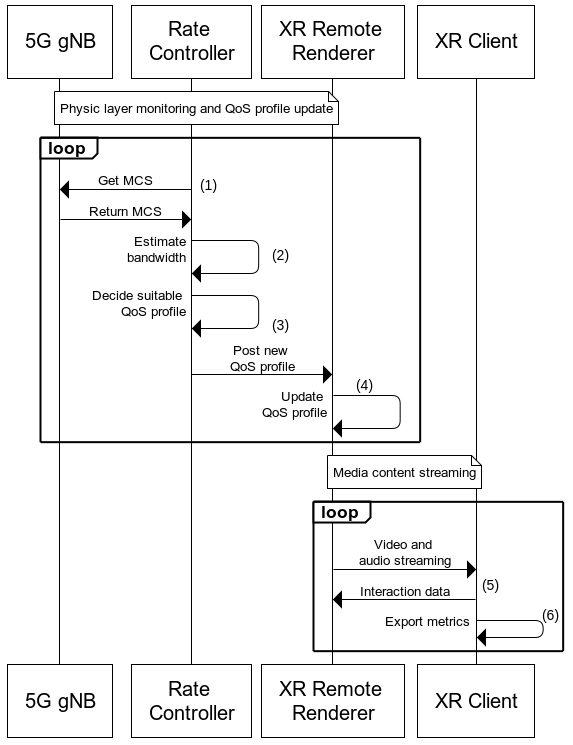}
     \caption{Sequence diagram.}
     \label{fig:sequence}
 \end{figure}

The communication between the various components of the system is illustrated in Figure \ref{fig:sequence}. The process of physic layer monitoring and the streaming quality profile updating is performed in a continuous loop. This cycle begins with the Rate Controller requesting the Modulation and Coding Scheme (MCS) index from the 5G gNB (1). The MCS index in a 5G network is a mechanism that determines how data is transmitted efficiently by combining modulation techniques and forward error correction coding rates. Modulation defines how digital bits are mapped to symbols, such as QPSK, 16-QAM, or 256-QAM, while the coding scheme adds redundancy to protect the data against errors. The MCS is dynamically adjusted based on the quality of the wireless channel; when conditions are favorable, higher-order modulation and lower redundancy allow for faster data rates, whereas in poor channel conditions, lower-order modulation with higher redundancy ensures more reliable communication, but reduce the achieved data rate. Upon receiving the MCS index, the Rate Controller estimates the wireless channel bandwidth using Equation \ref{deqn_ex1} (2). If necessary, it communicates with the XR Remote Renderer to adjust the streaming quality profile (3). The XR Remote Renderer then updates the content streaming configuration accordingly (4). Concurrently, the XR Client receives the rendered video and audio streams and sends interaction data back to the XR Remote Renderer (5). Additionally, the XR Client exports performance-related metrics for further analysis (6). To prevent an excessive number of requests for quality changes to the XR Remote Renderer, a limit of one change every three seconds has been applied. This mechanism helps avoid unnecessary switching in scenarios where the channel quality fluctuates slightly above and below the predefined thresholds. By imposing this restriction, the system ensures stability and prevents rapid oscillations between streaming quality profiles due to minor, short-term variations in network conditions.

Three predefined streaming quality profiles are configured for the XR service, each defined by resolution, frame rate and bitrate \cite{google_youtube_bitrate} as described in Table \ref{tab:transmission_profiles}. The maximum data rate calculated every second determines the selected streaming quality profile, enabling the adaptation of the XR rendering quality to the available network capacity in real-time. The data rate ranges have been assigned to each quality profile taking into account the overhead that the RTP protocol adds to the data transmission \cite{schulzrinne2003rtp}. 

\begin{table}[ht]
\renewcommand{\arraystretch}{1.9} % Increases the cell height
\caption{Streaming Quality Profiles for Content Delivery}
\centering
\begin{tabular}{|c|c|c|c|c|}
\hline
\textbf{Profile} & \textbf{Resolution} & \makecell{\textbf{Framerate} \\ \textbf{(fps)}} & \makecell{\textbf{Bitrate} \\ \textbf{(Mbps)}} & \makecell{\textbf{Data Rate} \\ \textbf{Condition (Mbps)}} \\ \hline
Q1               & 720p                & 60                           & 5                           & \textit{datarate} $<$ 8                            \\ \hline
Q2               & 1080p               & 60                           & 8                           & 8 $<$  \textit{datarate} $<$ 35                         \\ \hline
Q3               & 4K                  & 60                           & 35                          & 35$<$ \textit{datarate}                          \\ \hline
\end{tabular}
\label{tab:transmission_profiles}
\end{table}

\section{Experimental assessment}
\label{sec:evaluation}
\subsection{Testbed setup}
\label{sec:testbed}
To evaluate the theoretical data rate estimation and its impact on QoS in 5G networks, a testbed has been set up using the Amarisoft platform \cite{Amarisoft} to create a standalone 5G network. The Amarisoft software generates the 5G network, including both the core and radio access network (RAN) and provides detailed logs that include key network parameters. It has been configured with a specific network setup tailored to the requirements of the study. This configuration provides a controlled environment to evaluate the proposed methods effectively. The network parameters established in Amarisoft have been subsequently utilized to complete the bandwidth estimation equation. The key parameters of the testbed are listed in Table \ref{tab:formula_values}.

%  and work together to compose an auto-adaptive XR Remote Rendering system.

\begin{table}[ht]
\renewcommand{\arraystretch}{1.5} % Aumenta la altura de las celdas
\caption{Values used in the formula for Data Rate Calculation.}
\centering
\begin{tabular}{|l|c|}
\hline
% \textbf{Parameter}                      & \textbf{Value}                     \\ \hline
Number of Carriers (\(J\))              & 1                                  \\ \hline
Number of Layers (\(v_{\text{Layers}}\)) & 1                                 \\ \hline
Bandwidth (\(BW\))                       & 40 MHz                            \\ \hline
Subcarrier Spacing (\( \Delta f \))       & 30 kHz                           \\ \hline
Physical Resource Blocks (\(N_{\text{PRB}}\)) & 106 PRBs                         \\ \hline
Numerology (\( \mu \))                    & 1                                \\ \hline
% Modulation Order (\(Q_m\))              & 4 (16-QAM)                         \\ \hline
Scaling Factor (\(f\))                  & 1                               \\ \hline
% Maximum Code Rate (\(R_{\text{max}}\))  & \( \frac{948}{1024} \)             \\ \hline
% Physical Resource Blocks (\(N_{\text{PRB}}\)) & 106                             \\ \hline
Symbol Duration (\(T_s^\mu\))           & \(35.7143 \, \mu s\)               \\ \hline
Downlink Overhead (\(OH\))                       & 0.14     \\ \hline
Sampling Rate    ($f_s$)         & 1 second               \\ \hline
\end{tabular}
\label{tab:formula_values}
\end{table}

The Amarisoft testbed allocates 106 PRBs, derived from the configured bandwidth and subcarrier spacing \cite{3gpp_ts138_101_1_v15_2}. The Rate Controller processes the extracted MCS index, which determines the modulation order (\( Q_m \)), coding efficiency (\( f \)), and maximum code rate (\( R_{\text{max}} \)) as specified in \cite{3gpp_ts38_214_v15_2}. In this case, the maximum data rate is calculated for downlink (DL), using equation \ref{deqn_ex1}. 

The Rate Controller has been implemented using a Python script that communicates via WebSocket with the Amarisoft gNB to retrieve network quality metrics. After performing the estimation, it interacts with the XR Remote Renderer through an API, instructing it to adjust the streaming quality profile if necessary. The XR Remote Renderer has been implemented using the Unity framework \cite{Unity2025} and is responsible for generating the 3D scene and rendering it for transmission to the XR Client using the WebRTC protocol. The XR Client is a web-based player implemented in JavaScript, capable of running on any device with a browser and WebXR support. Regarding deployment, both the Rate Controller and XR Remote Renderer are hosted on an Edge Server with a GPU, while the XR Client is deployed on a laptop equipped with a 5G modem. This testbed setup provides a controlled environment to evaluate the relationship between the calculated data rate, the network configuration, and the adaptability of streaming quality for a remote rendering XR service.

\subsection{Results and discussion}
\label{sec:results}

To validate Equation \ref{deqn_ex1}, an experiment was conducted where the estimated data rate and actual throughput were measured simultaneously. The actual throughput was obtained using the \textit{iperf3} tool \cite{iperf3}, a widely used software for measuring network performance. The estimated data rate was calculated in real time using the parameters outlined in Section \ref{sec:testbed}, leveraging the data extracted from the Amarisoft logs.

\begin{figure}[t!]  
\centering 
\includegraphics[width=1
\linewidth]{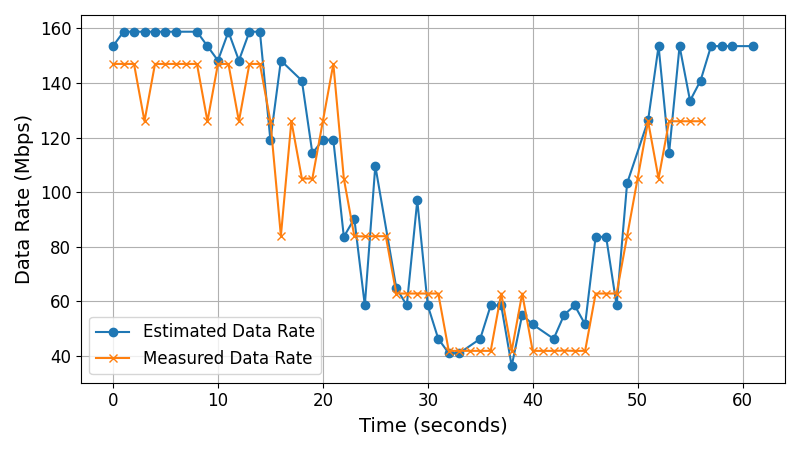}
\caption{Comparison between estimated data rate and  measured throughput over time.} 
\label{fig:comparison} 
\end{figure}

Figure \ref{fig:comparison} shows the results of this validation experiment. The graph compares the estimated maximum data rate, as calculated using the theoretical equation, with the actual data rate measured via \textit{iperf3} over a 60-second period. Both values were sampled at one-second intervals to ensure consistency in the comparison.

As shown in the figure, the estimated data rate closely follows the trend of the measured data rate, demonstrating the accuracy of the proposed equation under varying network conditions. While minor discrepancies can be observed, these are primarily attributed to dynamic network overheads, such as retransmissions and control signaling, which are not explicitly accounted for in the theoretical model.

This validation confirms that the proposed equation provides a reliable estimation of the available network capacity. Based on this, several experiments were conducted to observe the impact of the proposed auto-adaptive method on the XR service described earlier. To this end, QoS of a client consuming content generated by the XR remote renderer without any auto-adaptive quality method was compared to that obtained using the proposed auto-adaptive method.

To ensure a meaningful comparison of the impact of auto-adaptation, the evaluation was performed in a scenario where network capacity varied dynamically. Specifically, the test was designed so that the user equipment (UE), which runs the XR Client,  moved along a predefined trajectory, starting next to the transmission antenna, where network conditions were optimal, and gradually moving away until reaching a maximum distance of 15 meters. The UE then returned to its original position, completing the trajectory in 140 seconds. This mobility pattern introduced controlled variations in network capacity, allowing the auto-adaptive method to be evaluated under fluctuating conditions.

Initially, the remote renderer was launched without any auto-adaptation mechanism, and various QoS-related metrics were collected. In this scenario, the transmitted multimedia content was configured to the Q3 quality profile, as described in Table \ref{tab:transmission_profiles}. The same test was then repeated with the auto-adaptive method enabled, ensuring statistical robustness through multiple iterations.

\begin{figure}[t!]  
\centering 
\includegraphics[width=1
\linewidth]{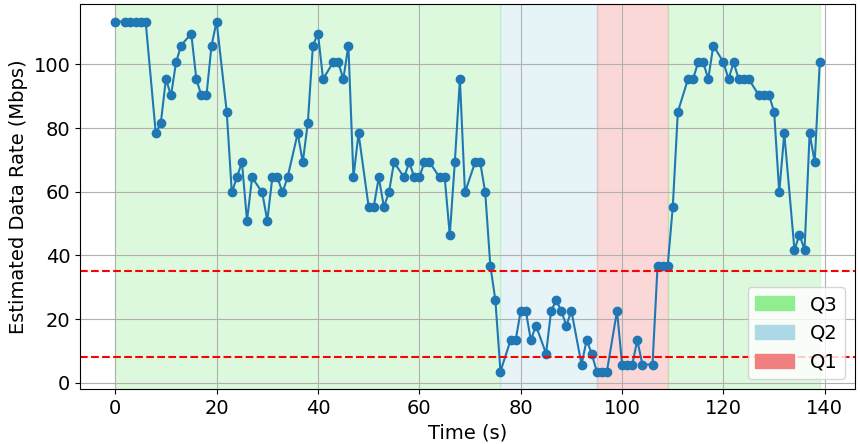}
\caption{Estimated data rate and current streaming profile.} 
\label{fig:datarate} 
\end{figure}

Figure \ref{fig:datarate} presents the estimated data rate during one of the tests conducted with auto-adaptation enabled. It also illustrates the streaming quality profile selected at each moment. In the graph, two horizontal lines can be seen, representing the data rate thresholds defined for selecting one streaming quality profile over another (8 Mbps and 36 Mbps).  The changes in the streaming quality profile correspond to the variability of the estimated bitrate, which ranges from 158.796 Mbps (MCS = 27) to 5.025 Mbps (MCS = 0).

The observed metrics included the number of switches (\textit{S$_{\textit{Nb}}$}), the number of freezes (\textit{F$_{\textit{Nb}}$}), the total freeze duration (\textit{F$_{\textit{tot}}$(ms)}), the average freeze duration (\textit{F$_{\textit{avg}}$(ms)}) and the average framerate (\textit{fps$_{\textit{avg}}$}), and are presented in Table \ref{tab:results}.

\begin{table}[t!]
\caption{Number of switches (\textit{S$_{\textit{Nb}}$}), number of freezes (\textit{F$_{\textit{Nb}}$}), total freeze duration (\textit{F$_{\textit{tot}}$(ms)}), average freeze duration (\textit{F$_{\textit{avg}}$(ms)}) and average framerate (\textit{fps$_{\textit{avg}}$}) for both No adaptation and Auto-adaptive Rate Control-enabled cases.}
\centering
\def\arraystretch{1.9}%  1 is the default, change whatever you need
\setlength\tabcolsep{2.5pt} % default value: 6pt
\label{tab:results}
\begin{tabular}{|c|c|c|c|c|c|}
\hline
\textbf{XR Remote Renderer} & \textbf{S$_{\textbf{Nb}}$} & \textbf{F$_{\textbf{Nb}}$} & \textbf{F$_{\textbf{tot}}$(ms)} & \textbf{F$_{\textbf{avg}}$(ms)}& \textbf{fps$_{\textbf{avg}}$}\\
\hline
No adaptation & 0 & 4 & 4760 & 1189 & 56.12  \\
\hline
Auto-adaptive Rate Control & 3 & 6 & 3930 & 786 & 56.51 \\
\hline
\end{tabular}
\end{table}

\begin{figure}[t!]  
\centering 
\includegraphics[width=1
\linewidth]{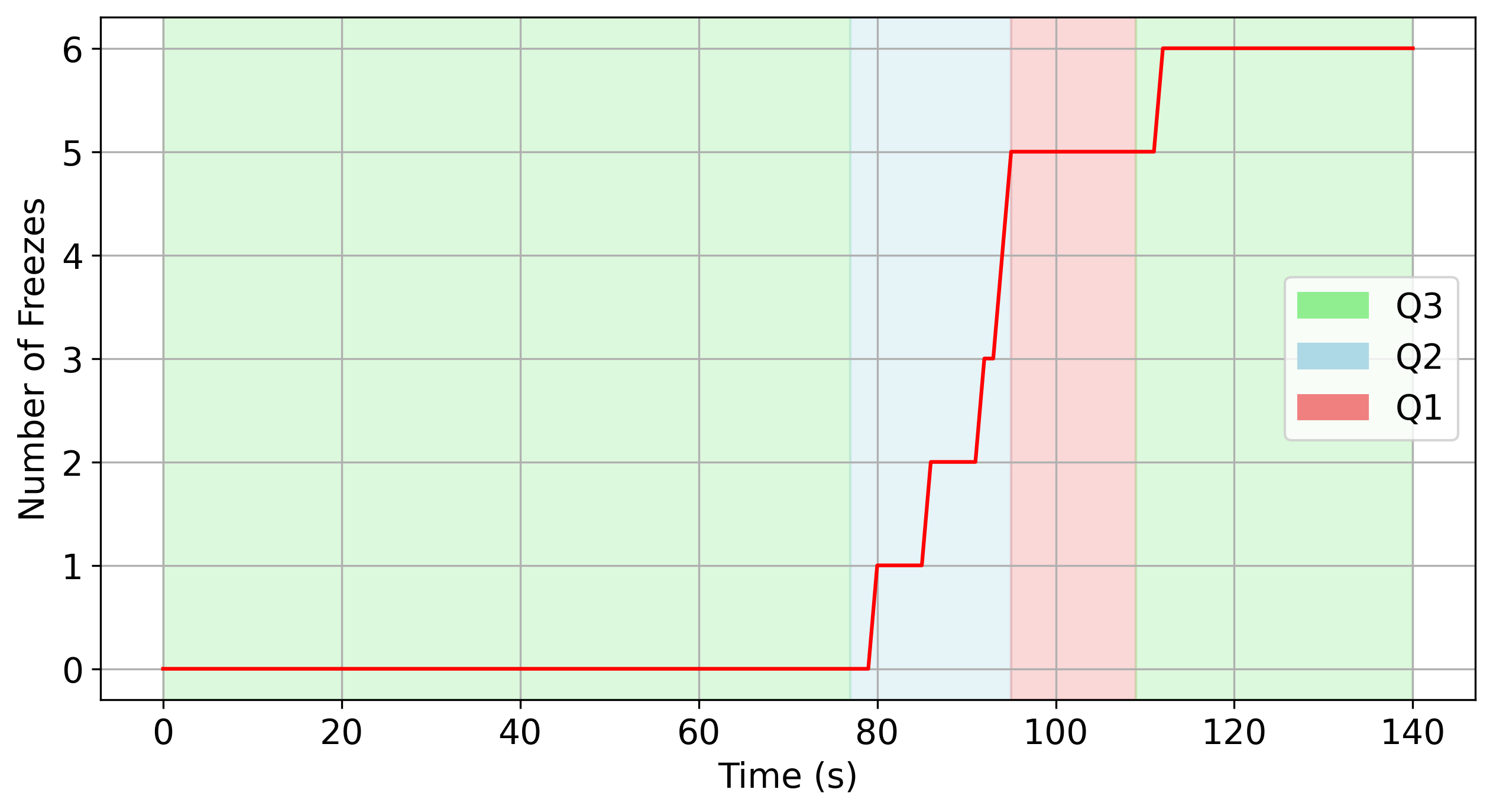}
\caption{Number of freezes and current streaming profile.} 
\label{fig:freezes} 
\end{figure}

In the case without adaptation, no changes in the streaming quality profile were observed, and four freezes occurred, resulting in a total freeze duration of 4760 ms. The average duration of each freeze was 1189 ms, while the overall average framerate throughout the test was 56.12 fps.

When applying the auto-adaptive method, three streaming quality profile switches were recorded during the test. In this case, six freezes were observed, with a total freeze duration of 3930 ms and an average freeze duration of 786 ms. The average framerate remained relatively stable at 56.51 fps. The variation in framerate is due to the freezes, which cause the framerate to drop.

Comparing the results of both tests, we observe that the auto-adaptive method led to a higher number of freezes. However, the total freeze duration and the average freeze duration were lower than in the non-adaptive case. Regarding the framerate, no significant differences were observed. To gain a deeper understanding of the obtained results, additional tests were conducted using the auto-adaptive system. These tests revealed that every time the remote renderer performs a streaming quality profile switch, the WebRTC video streaming detects a freeze. The experiment was repeated ten times, confirming that each streaming quality profile switch consistently causes a freeze with an average duration of 277 ms. This behavior is attributed to a limitation in the remote renderer’s implementation. Without this limitation, the number of freezes would be expected to be 3, and the total duration of freezes approximately 3099 ms. 

Figure \ref{fig:freezes} shows the freezes and changes in the streaming quality profile over time. The graph illustrates that between seconds 70 and 80, QoS begins to degrade as freezes start to happen, until around second 100, which corresponds to the moment when the XR Client is farthest from the base station, causing the channel quality to worsen. During this period, the remote renderer adjusts the video streaming quality profile, lowering it from Q3 to Q2 and later to Q1. Subsequently, the remote renderer detects optimal network capacity again and returns to the Q3 quality profile, restoring streaming smoothness.

Taking into account the limitations of the XR service used in the experiments, it is observed that, during 76.43\% of the time, the UE receives the content in the highest available quality. For the remaining time, the content is delivered at lower quality levels. This adaptation helps achieve a reduction of 17.4\% in the total freeze time. Furthermore, if the aforementioned limitations, particularly those related to freezes caused by changes in the quality of the stream, were not present, the total freeze time could be reduced by as much as 34.9\%. This demonstrates the potential for further optimization in the system, should these constraints be alleviated.

%By incorporating parameters such as modulation order, coding efficiency, and physical resource blocks, the model captures the core factors influencing throughput in a 5G network. These findings are critical for the subsequent application of this estimation in adaptive XR streaming, as discussed in the next subsection.

%The final version of this paper will include the results obtained from experiments conducted with a remote renderer for XR experiences, analyzing the impact of leveraging physical layer metrics on the system's QoS by comparing it with a setup that does not implement an auto-adaptation method.

\section{Conclusions and Future Work}
\label{sec:conclusions}
In this study, a theoretical model for estimating the maximum data rate of a 5G network has been utilized, and its potential application for enhancing the performance of a remote XR rendering service has been investigated. The findings demonstrate the positive impact of network bandwidth monitoring and estimation, combined with an auto-adaptive method to dynamically adjust the service's QoS according to the transmission channel capacity. The impact has been assessed by analyzing various metrics, including framerate, the number of freezes, and freeze duration. Furthermore, the results indicate that the observed performance is constrained by the implementation of the XR service employed in this study.

By bridging the gap between physical-layer network insights and XR application performance, this work lays the foundation for more resilient and adaptive XR systems. Future work will investigate how to leverage physical layer metrics to predict variations in network capacity, enabling auto-adaptation. This approach aims to further enhance system performance and quality of service.

\section*{Acknowledgment}
This research was supported by the SNS-JU Horizon Europe Research and Innovation programme, under Grant Agreement 101096838 for 6G-XR project. It has also been partially funded by the Spanish National Plan for Scientific and Technical Research and Innovation, under award number CER-20231018, project 6GDiferente and by the national project Grant PID2022-137329OB-C44 funded by MICIU/AEI/ 10.13039/501100011033 titled ”EnablIng Native-AI Secure deterministic 6G networks for hyPer-connected envIRonmEnts” (6G-INSPIRE).

\bibliographystyle{IEEEtran}
\bibliography{main.bib}

\end{document}